**Dispersion of speech aerosols in the context of physical distancing recommendations**

Vrishank Raghav, PhD [1]; Zu Puayen Tan, PhD [1]; Surya P. Bhatt, MD MSPH [2]

[1] *Department of Aerospace Engineering, Auburn University, Auburn, AL, 36849*

[2] *Division of Pulmonary, Allergy, and Critical Care Medicine, University of Alabama at Birmingham, Birmingham, AL 35294*

**Background**: Severe acute respiratory syndrome coronavirus 2 (SARS-CoV-2) can be transmitted by aerosols and droplets generated during cough as well as normal speech . The risk burden for contacts is affected by both physical distance and time of exposure. Current recommendations by the Centers for Disease Control and Prevention (CDC) define prolonged exposure operationally as 15 minutes of close exposure (within 6 feet) (3). The risk for contacts also depends on whether the source is symptomatic with cough or is asymptomatic wherein aerosols are generated by normal speech. Current understanding of normal speech generated aerosol-based transmission of SARS-CoV-2 is limited (1). We aimed to evaluate the dispersion extent of aerosol and gas plumes generated during vocalization of specific phones at varying sound intensity levels.

**Methods**: High-speed particle image velocimetry (PIV), a standard methodology to measure velocity fields (4), was used to quantify the dispersion of aerosol-laden gas-clouds generated during phonetic vocalization by a representative human subject (34 year old male of height 1.77m). Aerosol dispersion along the subject's mid-sagittal plane was characterized by flooding the indoor test environment (450 ft$^2$) with flow-tracing neutrally buoyant helium-filled soap bubbles, which were illuminated by a pulsed laser sheet (527nm, 18mJ/pulse at 300Hz) and recorded using two high-speed cameras (VEO640, 4MP at 300Hz, Vision Research) in frame straddle-mode (*Figure 1A and Supplement*). A commercially available software (DaVis 10, LaVision Inc) was used to process the raw data into velocity fields (measurement uncertainty ≤ 5%) for analysis. Based on settling speeds of aerosols (5), velocity field data obtained via PIV will accurately track the dispersion of smaller buoyant particles (≤ 20 microns), hypothesized to be responsible for transmission of viral infections (6). A microphone placed 1m from the subject was used to measure the sound intensity.

Aerosol propagation during vocalization of specific phones or cough (interrupted jets) occurs in two stages (7): rapid initial penetration followed by slow propagation. Penetration depth was quantified from the measured PIV data as the maximum horizontal distance traveled by the plume from the subject before reaching ambient velocity. Propagation beyond initial penetration was theoretically projected using classical scaling laws (7) for a transient non-buoyant puff (interrupted jet) $L(t) = C_1 t^n + C_2$. Here, $L(t)$ is the propagation distance as function of $t$ the propagation time, $n = 0.25$ for transient jets and puffs (7), $C_1$ and $C_2$ are constants that depends on the diameter of the mouth ($d \approx 1.5\ cm$) and the average velocity ($v_0$) at the mouth exit. Measured PIV data were used to determine $C_1$ and $C_2$, from which propagation distances were computed for time periods beyond the measured duration.

**Results**: Loud cough was associated with the highest initial penetration depth (up to 5.5 feet, *Figure 1A-B and Animation*). Among individual phones, the plosives (*/ti/* and */ta/*) had the highest initial penetration (up to 4 feet) followed by both the fricatives (*/si/* and */sa/*) and nasal sounds (up to 2.5 feet). The initial penetration depth of the phrase 'Stay Healthy' (comprised of most of the individual phones) was 2.9 feet. Among the plosives and cough, higher sound intensity levels were associated with higher penetration (*Figure 1B)*. The penetration depth of loud intensity plosives (4.1 feet) was comparable to normal (3.3 feet) and moderate (4.4 feet) cough. Based on theoretical projections of propagation time, the subsequent dispersion of the aerosol cloud was estimated to reach as far as 7-13 feet for plosives and 12-16 feet for cough within 2 minutes (*Figure 1C*).

**Discussion**: For the first time, this work quantifies the penetration distance of aerosol clouds generated during regular speech at different sound intensity levels. Our results indicate that the penetration distance was comparable between loud intensity speech (for example during singing, classroom lectures, parties etc.) and moderate intensity cough. This is particularly noteworthy because prior studies have shown that the size and quantity of aerosol generation are equivalent between some speech phones and coughing (8). Furthermore, based on theoretical aerosol propagation distance and time, we characterize potential exposure risk burden for contacts; around 2 minutes of exposure to an asymptomatic person as far as 7-13 feet apart during regular speech is likely to place contacts at risk of inhaling infectious aerosol. As such, physical distancing recommendations are likely sufficient to avoid incidental exposure by the initial penetration of the aerosol cloud, but insufficient for prolonged exposure to slow propagating aerosol clouds. While the concentration of virus particles in the aerosol cloud is bound to dilute with propagation distance and time, it is important to acknowledge the high virulence of SARS-CoV-2 (9) and the unknown infectious dose (10).

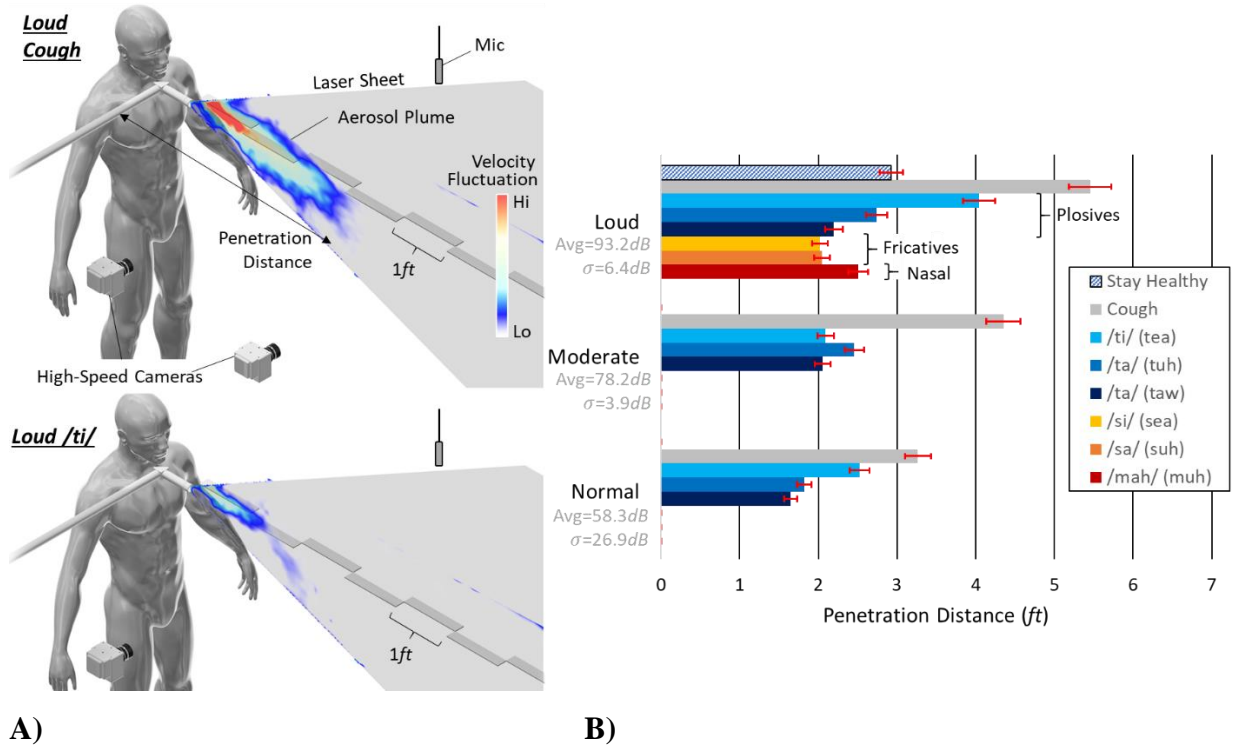

Figure 1: A) Aerosol plumes from loud cough and loud /ti/ vocalization. B) Penetration distances for different vocalizations and intensities. C) Theoretical propagation time of aerosol cloud/plume for normal and loud intensity cough and speech. Human model source - https://free3d.com/3d-model/male-base-mesh-6682.html

Our study has some limitations. Although our findings were generated by evaluation of a single subject, prior studies have reported minimal variance by age, gender, and BMI in aerosol generation during speech (11). Additionally, the propagation time and distance of the aerosol cloud beyond instantaneous penetration were estimated based on a theoretical model for a non-buoyant puff with no ambient air currents. This estimate represents an "average risk" scenario that is likely to be influenced (increase or decrease) by the directionality and magnitude of the prevailing indoor air currents.

In conclusion, given the undetermined infectious dose of SARS-CoV-2 and the expected cumulative aerosol dose in practical scenarios, such as several minutes of conversation, it is prudent to treat the six-foot recommendation as a conservative limit for short encounters. As wearing a mask can significantly alleviate aerosol dispersion (12), contact with an asymptomatic individual with SARS-CoV-2 for extended periods is likely unsafe, at any practical distancing limit, without a mask.

**Acknowledgment**: We gratefully acknowledge assistance from Lokesh Silwal for experimental setup and calibration. Research partially supported by National Institute of Biomedical Imaging and Bioengineering Trailblazer Award - 1R21EB027891. The acquisition of velocimetry instrumentation was partially supported under Army Research Office grant - W911NF-19-1-0124.

**References**

1. Liu Y, Ning Z, Chen Y, Guo M, Liu Y, Gali NK, Sun L, Duan Y, Cai J, Westerdahl DJN. Aerodynamic analysis of SARS-CoV-2 in two Wuhan hospitals. 2020: 1-6.
2. Rothe C, Schunk M, Sothmann P, Bretzel G, Froeschl G, Wallrauch C, Zimmer T, Thiel V, Janke C, Guggemos W. Transmission of 2019-nCoV infection from an asymptomatic contact in Germany. *New England Journal of Medicine* 2020; 382: 970-971.
3. Centers for Disease Control and Prevention. Public Health Recommendations for Community-Related Exposure. 2020.
4. Raffel M, Willert CE, Scarano F, Kähler CJ, Wereley ST, Kompenhans J. Particle image velocimetry: a practical guide. Springer; 2018.
5. Bourouiba L, Dehandschoewercker E, Bush JWJJoFM. Violent expiratory events: on coughing and sneezing. 2014; 745: 537-563.
6. Tellier R, Li Y, Cowling BJ, Tang JWJBid. Recognition of aerosol transmission of infectious agents: a commentary. 2019; 19: 101.
7. Sangras R, Kwon O, Faeth GJJHT. Self-preserving properties of unsteady round nonbuoyant turbulent starting jets and puffs in still fluids. 2002; 124: 460-469.
8. Chao CYH, Wan MP, Morawska L, Johnson GR, Ristovski Z, Hargreaves M, Mengersen K, Corbett S, Li Y, Xie X. Characterization of expiration air jets and droplet size distributions immediately at the mouth opening. *Journal of Aerosol Science* 2009; 40: 122-133.

9. Somsen GA, van Rijn C, Kooij S, Bem RA, Bonn DJTLRM. Small droplet aerosols in poorly ventilated spaces and SARS-CoV-2 transmission. 2020.
10. Schröder IJACH, Safety. COVID-19: A Risk Assessment Perspective. 2020.
11. Asadi S, Wexler AS, Cappa CD, Barreda S, Bouvier NM, Ristenpart WDJSr. Aerosol emission and superemission during human speech increase with voice loudness. 2019; 9: 1-10.
12. Tang JW, Liebner TJ, Craven BA, Settles GS. A schlieren optical study of the human cough with and without wearing masks for aerosol infection control. *Journal of the Royal Society Interface* 2009; 6: S727-S736.